\documentclass[5p]{elsarticle}

\usepackage{graphicx}
\usepackage{amssymb}

\usepackage{color}
\begin{document}

% ---------------- Start of some useful definitions ---------------------------

\def\kevc1{\ifmmode\mathrm{\ keV/{\mit c}}
          \else$\mathrm{\ keV/{\mit c}}$\fi}
\def\Mevc1{\ifmmode\mathrm{\ MeV/{\mit c}}
          \else$\mathrm{\ MeV/{\mit c}}$\fi}
\def\mevc1{\ifmmode\mathrm{\ MeV/{\mit c}}
          \else$\mathrm{\ MeV/{\mit c}}$\fi}
\def\gevc1{\ifmmode\mathrm{\ GeV/{\mit c}}
          \else$\mathrm{\ GeV/{\mit c}}$\fi}
\def\kevc2{\ifmmode\mathrm{\ keV/{\mit c}^2}
          \else$\mathrm{\ keV/{\mit c}^2}$\fi}
\def\Mevc2{\ifmmode\mathrm{\ MeV/{\mit c}^2}
          \else$\mathrm{\ MeV/{\mit c}^2}$\fi}
\def\Gevc2{\ifmmode\mathrm{\ GeV/{\mit c}^2}
          \else$\mathrm{\ GeV/{\mit c}^2}$\fi}
\def\Gev2c2{\ifmmode\mathrm{\ GeV^2/{\mit c}^2}
          \else$\mathrm{\ GeV^2/{\mit c}^2}$\fi}

\newcommand{\dcy}{\mbox{$ \rightarrow $}}

% quarks

\def\ubar{\ifmmode\mathrm{\overline {u}}
          \else$\mathrm{\overline{u}}$\fi}
\def\dbar{\ifmmode\mathrm{\overline {d}}
          \else$\mathrm{\overline{d}}$\fi}
\def\sbar{\ifmmode\mathrm{\overline {s}}
          \else$\mathrm{\overline{s}}$\fi}
\def\cbar{\ifmmode\mathrm{\overline {c}}
          \else$\mathrm{\overline{c}}$\fi}
\def\bbar{\ifmmode\mathrm{\overline {b}}
          \else$\mathrm{\overline{b}}$\fi}
\def\tbar{\ifmmode\mathrm{\overline {t}}
          \else$\mathrm{\overline{t}}$\fi}
\def\qbar{\ifmmode\mathrm{\overline {q}}
          \else$\mathrm{\overline{q}}$\fi}

\def\uq{\ifmmode\mathrm{u}
          \else$\mathrm{u}$\fi}
\def\dq{\ifmmode\mathrm{d}
          \else$\mathrm{d}$\fi}
\def\sq{\ifmmode\mathrm{s}
          \else$\mathrm{s}$\fi}
\def\cq{\ifmmode\mathrm{c}
          \else$\mathrm{c}$\fi}
\def\bq{\ifmmode\mathrm{b}
          \else$\mathrm{b}$\fi}
\def\tq{\ifmmode\mathrm{t}
          \else$\mathrm{t}$\fi}
\def\qq{\ifmmode\mathrm{q}
          \else$\mathrm{q}$\fi}

% particles

 \def\Pgg{\ifmmode\mathrm{\gamma}
          \else$\mathrm{\gamma}$\fi}
 \def\PW{\ifmmode\mathrm{W}
         \else$\mathrm{W }$\fi}
 \def\PWp{\ifmmode\mathrm{W^+}
          \else$\mathrm{W^+}$\fi}
 \def\PWpm{\ifmmode\mathrm{W^{\pm}}
          \else$\mathrm{W^{\pm}}$\fi}
 \def\PWm{\ifmmode\mathrm{W^-}
          \else$\mathrm{W^-}$\fi}
 \def\PZz{\ifmmode\mathrm{Z^0}
          \else$\mathrm{Z^0}$\fi}
 \def\PHz{\ifmmode\mathrm{H^0}
          \else$\mathrm{H^0}$\fi}
 \def\PHpm{\ifmmode\mathrm{H^{\pm}}
           \else$\mathrm{H^{\pm}}$\fi}
 \def\PWR{\ifmmode\mathrm{W_R}
          \else$\mathrm{W_R}$\fi}
 \def\PWpr{\ifmmode\mathrm{W^{\prime}}
           \else$\mathrm{W^{\prime}}$\fi}
 \def\PZLR{\ifmmode\mathrm{Z_{LR}}
           \else$\mathrm{Z_{LR}}$\fi}
 \def\PZgc{\ifmmode\mathrm{Z_{\chi}}
           \else$\mathrm{Z_{\chi}}$\fi}
 \def\PZgy{\ifmmode\mathrm{Z_{\psi}}
           \else$\mathrm{Z_{\psi}}$\fi}
 \def\PZge{\ifmmode\mathrm{Z_{\eta}}
           \else$\mathrm{Z_{\eta}}$\fi}
 \def\PZi{\ifmmode\mathrm{Z_1}
          \else$\mathrm{Z_1}$\fi}
 \def\PAz{\ifmmode\mathrm{A^0}
          \else$\mathrm{A^0}$\fi}
 \def\Pgne{\ifmmode\mathrm{\nu_{e}}
           \else$\mathrm{\nu_{e}}$\fi}
 \def\Pagne{\ifmmode\mathrm{\overline{\nu_{e}}}
            \else$\mathrm{\overline{\nu_{e}}}$\fi}
 \def\Pgngm{\ifmmode\mathrm{\nu_{\mu}}
            \else$\mathrm{\nu_{\mu}}$\fi}
 \def\Pagngm{\ifmmode\mathrm{\overline{\nu}_{\mu}}
             \else$\mathrm{\overline{\nu}_{\mu}}$\fi}
 \def\Pgngt{\ifmmode\mathrm{\nu_{\tau}}
            \else$\mathrm{\nu_{\tau}}$\fi}
 \def\Pagngt{\ifmmode\mathrm{\overline{\nu}_{\tau}}
             \else$\mathrm{\overline{\nu}_{\tau}}$\fi}
 \def\Pe{\ifmmode\mathrm{e}
         \else$\mathrm{e}$\fi}
 \def\Pep{\ifmmode\mathrm{e^+}
          \else$\mathrm{e^+}$\fi}
 \def\Pem{\ifmmode\mathrm{e^-}
          \else$\mathrm{e^-}$\fi}
 \def\Pgm{\ifmmode\mathrm{\mu}
          \else$\mathrm{\mu}$\fi}
 \def\Pgmm{\ifmmode\mathrm{\mu^-}
           \else$\mathrm{\mu^-}$\fi}
 \def\Pgmp{\ifmmode\mathrm{\mu^+}
           \else$\mathrm{\mu^+}$\fi}
 \def\Pgt{\ifmmode\mathrm{\tau}
          \else$\mathrm{\tau}$\fi}
 \def\PLpm{\ifmmode\mathrm{L^{\pm}}
           \else$\mathrm{L^{\pm}}$\fi}
 \def\PLz{\ifmmode\mathrm{L^0}
          \else$\mathrm{L^0}$\fi}
 \def\PEz{\ifmmode\mathrm{E^0}
          \else$\mathrm{E^0}$\fi}
 \def\Pgp{\ifmmode\mathrm{\pi}
          \else$\mathrm{\pi }$\fi}
 \def\Pgpm{\ifmmode\mathrm{\pi^-}
           \else$\mathrm{\pi^-}$\fi}
 \def\Pgpp{\ifmmode\mathrm{\pi^+}
           \else$\mathrm{\pi^+}$\fi}
 \def\Pgppm{\ifmmode\mathrm{\pi^{\pm }}
            \else$\mathrm{\pi^{\pm }}$\fi}
 \def\Pgpz{\ifmmode\mathrm{\pi^0}
           \else$\mathrm{\pi^0 }$\fi}
 \def\Pgh{\ifmmode\mathrm{\eta}
          \else$\mathrm{\eta }$\fi}
 \def\Pgr{\ifmmode\mathrm{\rho(770)}
          \else$\mathrm{\rho(770)}$\fi}
 \def\Pgo{\ifmmode\mathrm{\omega(783)}
          \else$\mathrm{\omega(783)}$\fi}
 \def\Pghpr{\ifmmode\mathrm{\eta^{\prime}(958)}
            \else$\mathrm{\eta^{\prime}(958)}$\fi}
 \def\Pfz{\ifmmode\mathrm{f_0(980)}
          \else$\mathrm{f_0(980)}$\fi}
 \def\Paz{\ifmmode\mathrm{a_0(980)}
          \else$\mathrm{a_0(980)}$\fi}
 \def\Pgf{\ifmmode\mathrm{\phi(1020)}
          \else$\mathrm{\phi(1020)}$\fi}
 \def\Phia{\ifmmode\mathrm{h_1(1170)}
           \else$\mathrm{h_1(1170)}$\fi}
 \def\Pbi{\ifmmode\mathrm{b_1(1235)}
          \else$\mathrm{b_1(1235)}$\fi}
 \def\Pai{\ifmmode\mathrm{a_1(1260)}
          \else$\mathrm{a_1(1260)}$\fi}
 \def\Pfii{\ifmmode\mathrm{f_2(1270)}
           \else$\mathrm{f_2(1270)}$\fi}
 \def\Pfi{\ifmmode\mathrm{f_1(1285)}
          \else$\mathrm{f_1(1285)}$\fi}
 \def\Pgha{\ifmmode\mathrm{\eta(1295)}
           \else$\mathrm{\eta(1295)}$\fi}
 \def\Pgpa{\ifmmode\mathrm{\pi(1300)}
           \else$\mathrm{\pi(1300)}$\fi}
 \def\Paii{\ifmmode\mathrm{a_2(1320)}
           \else$\mathrm{a_2(1320)}$\fi}
 \def\Pgoa{\ifmmode\mathrm{\omega(1390)}
           \else$\mathrm{\omega(1390)}$\fi}
 \def\Pfza{\ifmmode\mathrm{f_0(1400)}
           \else$\mathrm{f_0(1400)}$\fi}
 \def\Pfia{\ifmmode\mathrm{f_1 (1390)}
           \else$\mathrm{f_1 (1390)}$\fi}
 \def\Pghb{\ifmmode\mathrm{\eta(1440)}
           \else$\mathrm{\eta(1440)}$\fi}
 \def\Pgra{\ifmmode\mathrm{\rho(1450)}
           \else$\mathrm{\rho(1450)}$\fi}
 \def\Pfib{\ifmmode\mathrm{f_1(1510)}
           \else$\mathrm{f_1(1510)}$\fi}
 \def\Pfiipr{\ifmmode\mathrm{f^{\prime}_2(1525)}
             \else$\mathrm{f^{\prime}_2(1525)}$\fi}
 \def\Pfzb{\ifmmode\mathrm{f_0(1590)}
           \else$\mathrm{f_0(1590)}$\fi}
 \def\Pgob{\ifmmode\mathrm{\omega(1600)}
           \else$\mathrm{\omega(1600)}$\fi}
 \def\Pgoiii{\ifmmode\mathrm{\omega_3(1670)}
             \else$\mathrm{\omega_3(1670)}$\fi}
 \def\Pgpii{\ifmmode\mathrm{\pi_2(1670)}
            \else$\mathrm{\pi_2(1670)}$\fi}
 \def\Pgfa{\ifmmode\mathrm{\phi(1680)}
           \else$\mathrm{\phi(1680)}$\fi}
 \def\Pgriii{\ifmmode\mathrm{\rho_3(1690)}
             \else$\mathrm{\rho_3(1690)}$\fi}
 \def\Pgrb{\ifmmode\mathrm{\rho(1700)}
           \else$\mathrm{\rho(1700)}$\fi}
 \def\Pfiia{\ifmmode\mathrm{f_2(1720)}
            \else$\mathrm{f_2(1720)}$\fi}
 \def\Pgfiii{\ifmmode\mathrm{\phi_3(1850)}
             \else$\mathrm{\phi_3(1850)}$\fi}
 \def\Pfiib{\ifmmode\mathrm{f_2(2010)}
            \else$\mathrm{f_2(2010)}$\fi}
 \def\Pfiv{\ifmmode\mathrm{f_4(2050)}
           \else$\mathrm{f_4(2050)}$\fi}
 \def\Pfiic{\ifmmode\mathrm{f_2(2300)}
            \else$\mathrm{f_2(2300)}$\fi}
 \def\Pfiid{\ifmmode\mathrm{f_2(2340)}
            \else$\mathrm{f_2(2340)}$\fi}
 \def\PK{\ifmmode\mathrm{K}
         \else$\mathrm{K}$\fi}
 \def\PKpm{\ifmmode\mathrm{K^{\pm}}
           \else$\mathrm{K^{\pm}}$\fi}
 \def\PKp{\ifmmode\mathrm{K^+}
          \else$\mathrm{K^+}$\fi}
 \def\PKm{\ifmmode\mathrm{K^-}
          \else$\mathrm{K^-}$\fi}
 \def\PKz{\ifmmode\mathrm{K^0}
          \else$\mathrm{K^0}$\fi}
 \def\PaKz{\ifmmode\mathrm{\overline{K^0}}
           \else$\mathrm{\overline{K^0}}$\fi}
 \def\PKgmiii{\ifmmode\mathrm{K_{\mu 3}}
              \else$\mathrm{K_{\mu 3}}$\fi}
 \def\PKeiii{\ifmmode\mathrm{K_{\rm e3}}
             \else$\mathrm{K_{\rm e3}}$\fi}
 \def\PKzS{\ifmmode\mathrm{K^0_{\rm S}}
           \else$\mathrm{K^0_{\rm S}}$\fi}
 \def\PKzL{\ifmmode\mathrm{K^0_{\rm L}}
           \else$\mathrm{K^0_{\rm L}}$\fi}
 \def\PKzgmiii{\ifmmode\mathrm{K^0_{\mu 3}}
               \else$\mathrm{K^0_{\mu 3}}$\fi}
 \def\PKzeiii{\ifmmode\mathrm{K^0_{{\rm e}3}}
              \else$\mathrm{K^0_{{\rm e}3}}$\fi}
 \def\PKst{\ifmmode\mathrm{K^{\ast}(892)}
           \else$\mathrm{K^{\ast}(892)}$\fi}
 \def\PKi{\ifmmode\mathrm{K_1(1270)}
          \else$\mathrm{K_1(1270)}$\fi}
 \def\PKsta{\ifmmode\mathrm{K^{\ast}(1370)}
            \else$\mathrm{K^{\ast}(1370)}$\fi}
 \def\PKia{\ifmmode\mathrm{K_1(1400)}
           \else$\mathrm{K_1(1400)}$\fi}
 \def\PKstz{\ifmmode\mathrm{K^{\ast}_0(1430)}
            \else$\mathrm{K^{\ast}_0(1430)}$\fi}
 \def\PKstii{\ifmmode\mathrm{K^{\ast}_2(1430)}
             \else$\mathrm{K^{\ast}_2(1430)}$\fi}
 \def\PKstb{\ifmmode\mathrm{K^{\ast}(1680)}
            \else$\mathrm{K^{\ast}(1680)}$\fi}
 \def\PKii{\ifmmode\mathrm{K_2(1770)}
           \else$\mathrm{K_2(1770)}$\fi}
 \def\PKstiii{\ifmmode\mathrm{K^{\ast}_3(1780)}
              \else$\mathrm{K^{\ast}_3(1780)}$\fi}
 \def\PKstiv{\ifmmode\mathrm{K^{\ast}_4(2045)}
             \else$\mathrm{K^{\ast}_4(2045)}$\fi}
 \def\PD{\ifmmode\mathrm{D}
           \else$\mathrm{D}$\fi}
 \def\PaD{\ifmmode\mathrm{\overline{ D}}
          \else${\mathrm{\overline D}}$\fi}
 \def\PDpm{\ifmmode\mathrm{D^{\pm}}
           \else$\mathrm{D^{\pm}}$\fi}
 \def\PDm{\ifmmode\mathrm{D^-}
          \else$\mathrm{D^-}$\fi}
 \def\PDp{\ifmmode\mathrm{D^+}
          \else$\mathrm{D^+}$\fi}
 \def\PDz{\ifmmode\mathrm{D^0}
          \else$\mathrm{D^0}$\fi}
 \def\PaDz{\ifmmode\mathrm{\overline{D^0}}
           \else$\mathrm{\overline{D^0}}$\fi}
 \def\PDstpm{\ifmmode{\mathrm{D}^{\ast}(2010)^{\pm}}
             \else$\mathrm{D}^{\ast}(2010)^{\pm}$\fi}
 \def\PDstp{\ifmmode{\mathrm{D}^{\ast+}}
             \else$\mathrm{D}^{\ast+}$\fi}
 \def\PDst{\ifmmode{\mathrm{D}^{\ast}}
             \else$\mathrm{D}^{\ast}$\fi}
 \def\PDstz{\ifmmode{\mathrm{D}^{\ast}(2010)^0}
            \else$\mathrm{D}^{\ast}(2010)^0$\fi}
 \def\PDiz{\ifmmode{\mathrm{D}_{1}(2420)^0}
           \else$\mathrm{D}_{1}(2420)^0$\fi}
 \def\PDstiiz{\ifmmode{\mathrm{D}^{\ast}_{2}(2460)^0}
              \else$\mathrm{D}^{\ast}_{2}(2460)^0$\fi}
 \def\PsDp{\ifmmode\mathrm{D_{s}^+}
           \else$\mathrm{D_{s}^+}$\fi}
 \def\PsDm{\ifmmode\mathrm{D_{s}^-}
           \else$\mathrm{D_{s}^-}$\fi}
 \def\PsDpm{\ifmmode\mathrm{D_{s}^{\pm}}
           \else$\mathrm{D_{s}^{\pm}}$\fi}
 \def\PsDst{\ifmmode\mathrm{D_{s}^{\ast}}
            \else$\mathrm{D_{s}^{\ast}}$\fi}
 \def\PsDipm{\ifmmode\mathrm{D_{s1}(2536)^{\pm}}
           \else$\mathrm{D_{s1}(2536)^{\pm}}$\fi}
 \def\PB{\ifmmode{\mathrm{B}}
          \else$\mathrm{B}$\fi}
 \def\PBp{\ifmmode{\mathrm{B}^{+}}
           \else$\mathrm{B}^{+}$\fi}
 \def\PBm{\ifmmode{\mathrm{B}^{-}}
           \else$\mathrm{B}^{-}$\fi}
 \def\PBpm{\ifmmode{\mathrm{B}^{\pm}}
            \else$\mathrm{B}^{\pm}$\fi}
 \def\PBz{\ifmmode{\mathrm{B}^0}
           \else$\mathrm{B}^0$\fi}
 \def\PbgL{\ifmmode{\mathrm{\Lambda}_b}
           \else$\mathrm{\Lambda}_b$\fi}
 \def\Pcgh{\ifmmode\mathrm{{\eta}_{c}(1S)}
           \else$\mathrm{{\eta}_{c}(1S)}$\fi}
 \def\PJgyy{\ifmmode\mathrm{J /\psi}
           \else$\mathrm{J /\psi}$\fi}
 \def\PJgy{\ifmmode\mathrm{J /\psi(1S)}
           \else$\mathrm{J /\psi(1S)}$\fi}
 \def\Pcgcz{\ifmmode\mathrm{{\chi}_{c0}(1P)}
            \else$\mathrm{{\chi}_{c0}(1P)}$\fi}
 \def\Pcgci{\ifmmode\mathrm{{\chi}_{c1}(1P)}
            \else$\mathrm{{\chi}_{c1}(1P)}$\fi}
 \def\Pcgcii{\ifmmode\mathrm{{\chi}_{c2}(1P)}
             \else$\mathrm{{\chi}_{c2}(1P)}$\fi}
 \def\Pgy{\ifmmode\mathrm{\psi(2S)}
          \else$\mathrm{\psi(2S)}$\fi}
 \def\Pgya{\ifmmode\mathrm{\psi(3770)}
           \else$\mathrm{\psi(3770)}$\fi}
 \def\Pgyb{\ifmmode\mathrm{\psi(4040)}
           \else$\mathrm{\psi(4040)}$\fi}
 \def\Pgyc{\ifmmode\mathrm{\psi(4160)}
           \else$\mathrm{\psi(4160)}$\fi}
 \def\Pgyd{\ifmmode\mathrm{\psi(4415)}
           \else$\mathrm{\psi(4415)}$\fi}
 \def\PgU{\ifmmode\mathrm{\Upsilon(1S)}
          \else$\mathrm{\Upsilon(1S)}$\fi}
 \def\Pbgcz{\ifmmode\mathrm{{\chi}_{b0}(1P)}
            \else$\mathrm{{\chi}_{b0}(1P)}$\fi}
 \def\Pbgci{\ifmmode\mathrm{{\chi}_{b1}(1P)}
            \else$\mathrm{{\chi}_{b1}(1P)}$\fi}
 \def\Pbgcii{\ifmmode\mathrm{{\chi}_{b2}(1P)}
             \else$\mathrm{{\chi}_{b2}(1P)}$\fi}
 \def\PgUa{\ifmmode\mathrm{\Upsilon(2S)}
           \else$\mathrm{\Upsilon(2S)}$\fi}
 \def\Pbgcza{\ifmmode\mathrm{{\chi}_{b0}(2P)}
             \else$\mathrm{{\chi}_{b0}(2P)}$\fi}
 \def\Pbgcia{\ifmmode\mathrm{{\chi}_{b1}(2P)}
             \else$\mathrm{{\chi}_{b1}(2P)}$\fi}
 \def\Pbgciia{\ifmmode\mathrm{{\chi}_{b2}(2P)}
              \else$\mathrm{{\chi}_{b2}(2P)}$\fi}
 \def\PgUb{\ifmmode\mathrm{\Upsilon(3S)}
           \else$\mathrm{\Upsilon(3S)}$\fi}
 \def\PgUc{\ifmmode\mathrm{\Upsilon(4S)}
           \else$\mathrm{\Upsilon(4S)}$\fi}
 \def\PgUd{\ifmmode\mathrm{\Upsilon(10860)}
           \else$\mathrm{\Upsilon(10860)}$\fi}
 \def\PgUe{\ifmmode\mathrm{\Upsilon(11020)}
           \else$\mathrm{\Upsilon(11020)}$\fi}
 \def\Pp{\ifmmode\mathrm{p}
         \else$\mathrm{p}$\fi}
 \def\Pap{\ifmmode\mathrm{\overline{p}}
         \else$\mathrm{\overline{p}}$\fi}
 \def\Pn{\ifmmode\mathrm{n}
         \else$\mathrm{n}$\fi}
 \def\PNa{\ifmmode\mathrm{N(1440)P_{11}}
          \else$\mathrm{N(1440)P_{11}}$\fi}
 \def\PNb{\ifmmode\mathrm{N(1520)D_{13}}
          \else$\mathrm{N(1520)D_{13}}$\fi}
 \def\PNc{\ifmmode\mathrm{N(1535)S_{11}}
          \else$\mathrm{N(1535)S_{11}}$\fi}
 \def\PNd{\ifmmode\mathrm{N(1650)S_{11}}
          \else$\mathrm{N(1650)S_{11}}$\fi}
 \def\PNe{\ifmmode\mathrm{N(1675)D_{15}}
          \else$\mathrm{N(1675)D_{15}}$\fi}
 \def\PNf{\ifmmode\mathrm{N(1680)F_{15}}
          \else$\mathrm{N(1680)F_{15}}$\fi}
 \def\PNg{\ifmmode\mathrm{N(1700)D_{13}}
          \else$\mathrm{N(1700)D_{13}}$\fi}
 \def\PNh{\ifmmode\mathrm{N(1710)P_{11}}
          \else$\mathrm{N(1710)P_{11}}$\fi}
 \def\PNi{\ifmmode\mathrm{N(1720)P_{13}}
          \else$\mathrm{N(1720)P_{13}}$\fi}
 \def\PNj{\ifmmode\mathrm{N(2190)G_{17}}
          \else$\mathrm{N(2190)G_{17}}$\fi}
 \def\PNk{\ifmmode\mathrm{N(2220)H_{19}}
          \else$\mathrm{N(2220)H_{19}}$\fi}
 \def\PNl{\ifmmode\mathrm{N(2250)G_{19}}
          \else$\mathrm{N(2250)G_{19}}$\fi}
 \def\PNm{\ifmmode\mathrm{N(2600)I_{1,11}}
          \else$\mathrm{N(2600)I_{1,11}}$\fi}
 \def\PgDa{\ifmmode\mathrm{\Delta(1232)P_{33}}
           \else$\mathrm{\Delta(1232)P_{33}}$\fi}
 \def\PgDb{\ifmmode\mathrm{\Delta(1620)S_{31}}
           \else$\mathrm{\Delta(1620)S_{31}}$\fi}
 \def\PgDc{\ifmmode\mathrm{\Delta(1700)D_{33}}
           \else$\mathrm{\Delta(1700)D_{33}}$\fi}
 \def\PgDd{\ifmmode\mathrm{\Delta(1900)S_{31}}
           \else$\mathrm{\Delta(1900)S_{31}}$\fi}
 \def\PgDe{\ifmmode\mathrm{\Delta(1905)F_{35}}
           \else$\mathrm{\Delta(1905)F_{35}}$\fi}
 \def\PgDf{\ifmmode\mathrm{\Delta(1910)P_{31}}
           \else$\mathrm{\Delta(1910)P_{31}}$\fi}
 \def\PgDh{\ifmmode\mathrm{\Delta(1920)P_{33}}
           \else$\mathrm{\Delta(1920)P_{33}}$\fi}
 \def\PgDi{\ifmmode\mathrm{\Delta(1930)D_{35}}
           \else$\mathrm{\Delta(1930)D_{35}}$\fi}
 \def\PgDj{\ifmmode\mathrm{\Delta(1950)F_{37}}
           \else$\mathrm{\Delta(1950)F_{37}}$\fi}
 \def\PgDk{\ifmmode\mathrm{\Delta(2420)H_{3,11}}
           \else$\mathrm{\Delta(2420)H_{3,11}}$\fi}
 \def\PgDpp{\ifmmode\mathrm{\Delta^{++}}
           \else$\mathrm{\Delta^{++}}$\fi}
 \def\PgL{\ifmmode\mathrm{\Lambda}
          \else$\mathrm{\Lambda}$\fi}
 \def\PagL{\ifmmode\mathrm{\overline{\Lambda}}
            \else$\mathrm{\overline{\Lambda}}$\fi}
 \def\PgLa{\ifmmode\mathrm{\Lambda(1405) S_{01}}
           \else$\mathrm{\Lambda(1405) S_{01}}$\fi}
 \def\PgLb{\ifmmode\mathrm{\Lambda(1520) D_{03}}
           \else$\mathrm{\Lambda(1520) D_{03}}$\fi}
 \def\PgLc{\ifmmode\mathrm{\Lambda(1600) P_{01}}
           \else$\mathrm{\Lambda(1600) P_{01}}$\fi}
 \def\PgLd{\ifmmode\mathrm{\Lambda(1670) S_{01}}
           \else$\mathrm{\Lambda(1670) S_{01}}$\fi}
 \def\PgLe{\ifmmode\mathrm{\Lambda(1690) D_{03}}
           \else$\mathrm{\Lambda(1690) D_{03}}$\fi}
 \def\PgLf{\ifmmode\mathrm{\Lambda(1800) S_{01}}
           \else$\mathrm{\Lambda(1800) S_{01}}$\fi}
 \def\PgLg{\ifmmode\mathrm{\Lambda(1810) P_{01}}
           \else$\mathrm{\Lambda(1810) P_{01}}$\fi}
 \def\PgLh{\ifmmode\mathrm{\Lambda(1820) F_{05}}
           \else$\mathrm{\Lambda(1820) F_{05}}$\fi}
 \def\PgLi{\ifmmode\mathrm{\Lambda(1830) D_{05}}
           \else$\mathrm{\Lambda(1830) D_{05}}$\fi}
 \def\PgLj{\ifmmode\mathrm{\Lambda(1890) P_{03}}
           \else$\mathrm{\Lambda(1890) P_{03}}$\fi}
 \def\PgLk{\ifmmode\mathrm{\Lambda(2100) G_{07}}
           \else$\mathrm{\Lambda(2100) G_{07}}$\fi}
 \def\PgLl{\ifmmode\mathrm{\Lambda(2110) F_{05}}
           \else$\mathrm{\Lambda(2110) F_{05}}$\fi}
 \def\PgLm{\ifmmode\mathrm{\Lambda(2350) H_{09}}
           \else$\mathrm{\Lambda(2350) H_{09}}$\fi}
 \def\PgS{\ifmmode{\rm \Sigma}
           \else${\rm \Sigma}$\fi}
 \def\PgSp{\ifmmode\mathrm{\Sigma^+}
           \else$\mathrm{\Sigma^+}$\fi}
 \def\PgSz{\ifmmode\mathrm{\Sigma^0}
           \else$\mathrm{\Sigma^0}$\fi}
 \def\PgSm{\ifmmode\mathrm{\Sigma^-}
           \else$\mathrm{\Sigma^-}$\fi}
 \def\PgSpm{\ifmmode\mathrm{\Sigma^{\pm}}
           \else$\mathrm{\Sigma^{\pm}}$\fi}
 \def\PgSa{\ifmmode\mathrm{\Sigma(1385) P_{13}}
           \else$\mathrm{\Sigma(1385) P_{13}}$\fi}
 \def\PgSb{\ifmmode\mathrm{\Sigma(1660) P_{11}}
           \else$\mathrm{\Sigma(1660) P_{11}}$\fi}
 \def\PgSc{\ifmmode\mathrm{\Sigma(1670) D_{13}}
           \else$\mathrm{\Sigma(1670) D_{13}}$\fi}
 \def\PgSd{\ifmmode\mathrm{\Sigma(1750) S_{11}}
           \else$\mathrm{\Sigma(1750) S_{11}}$\fi}
 \def\PgSe{\ifmmode\mathrm{\Sigma(1775) D_{15}}
           \else$\mathrm{\Sigma(1775) D_{15}}$\fi}
 \def\PgSf{\ifmmode\mathrm{\Sigma(1915) F_{15}}
           \else$\mathrm{\Sigma(1915) F_{15}}$\fi}
 \def\PgSg{\ifmmode\mathrm{\Sigma(1940) D_{13}}
           \else$\mathrm{\Sigma(1940) D_{13}}$\fi}
 \def\PgSh{\ifmmode\mathrm{\Sigma(2030) F_{17}}
           \else$\mathrm{\Sigma(2030) F_{17}}$\fi}
 \def\PgSi{\ifmmode\mathrm{\Sigma(2050)}
           \else$\mathrm{\Sigma(2050)}$\fi}

 \def\PgXz{\ifmmode\mathrm{\Xi^0}
           \else$\mathrm{\Xi^0}$\fi}
 \def\PgXm{\ifmmode\mathrm{\Xi^-}
           \else$\mathrm{\Xi^-}$\fi}
 \def\PgXa{\ifmmode\mathrm{\Xi(1530)}
           \else$\mathrm{\Xi(1530)}$\fi}
 \def\PgXas{\ifmmode\mathrm{\Xi(1530)P_{13}}
           \else$\mathrm{\Xi(1530)P_{13}}$\fi}
 \def\PgXb{\ifmmode\mathrm{\Xi(1690)}
           \else$\mathrm{\Xi(1690)}$\fi}
 \def\PgXbb{\ifmmode\mathrm{\Xi(1620)}
           \else$\mathrm{\Xi(1620)}$\fi}
 \def\PgXc{\ifmmode\mathrm{\Xi(1820)D_{13}}
           \else$\mathrm{\Xi(1820)D_{13}}$\fi}
 \def\PgXcs{\ifmmode\mathrm{\Xi(1820)}
           \else$\mathrm{\Xi(1820)}$\fi}
 \def\PgXd{\ifmmode\mathrm{\Xi(1950)}
           \else$\mathrm{\Xi(1950)}$\fi}
 \def\PgXe{\ifmmode\mathrm{\Xi(2030)}
           \else$\mathrm{\Xi(2030)}$\fi}
 \def\PgOm{\ifmmode\mathrm{\Omega^-}
           \else$\mathrm{\Omega^-}$\fi}
 \def\PgO{\ifmmode\mathrm{\Omega}
           \else$\mathrm{\Omega}$\fi}
 \def\PgOma{\ifmmode\mathrm{\Omega(2250)^-}
            \else$\mathrm{\Omega(2250)^-}$\fi}
 \def\PcgL{\ifmmode\mathrm{\Lambda_c}
            \else$\mathrm{\Lambda_c}$\fi}
 \def\PacgL{\ifmmode\mathrm{\overline{\Lambda}_c}
            \else$\mathrm{\overline{\Lambda}_c}$\fi}
 \def\PcgLp{\ifmmode\mathrm{\Lambda_c^+}
            \else$\mathrm{\Lambda_c^+}$\fi}
 \def\PcgLm{\ifmmode{\rm \Lambda_c^-}
            \else${\rm \Lambda_c^-}$\fi}
 \def\PcgX{\ifmmode\mathrm{\Xi_c}
            \else$\mathrm{\Xi_c}$\fi}
 \def\PcgXz{\ifmmode\mathrm{\Xi_c^0}
            \else$\mathrm{\Xi_c^0}$\fi}
 \def\PcgXp{\ifmmode\mathrm{\Xi_c^+}
            \else$\mathrm{\Xi_c^+}$\fi}
 \def\PcgS{\ifmmode\mathrm{\Sigma_c}
           \else$\mathrm{\Sigma_c}$\fi}
 \def\PcgSz{\ifmmode\mathrm{\Sigma_c^0}
           \else$\mathrm{\Sigma_c^0}$\fi}
 \def\PcgSp{\ifmmode\mathrm{\Sigma_c^+}
           \else$\mathrm{\Sigma_c^+}$\fi}
 \def\PcgSpp{\ifmmode\mathrm{\Sigma_c^{++}}
           \else$\mathrm{\Sigma_c^{++}}$\fi}
 \def\PcgO{\ifmmode{\mathrm \Omega_c}
           \else${\mathrm \Omega_c}$\fi}
 \def\PcgOz{\ifmmode{\mathrm \Omega_c^{0}}
           \else${\mathrm \Omega_c^{0}}$\fi}
 \def\PSgg{\ifmmode\mathrm{\tilde{\gamma}}
           \else$\mathrm{\tilde{\gamma}}$\fi}
 \def\PSgxz{\ifmmode\mathrm{\tilde{\chi}^0_i}
            \else$\mathrm{\tilde{\chi}^0_i}$\fi}
 \def\PSZz{\ifmmode\mathrm{\tilde{Z}^0}
           \else$\mathrm{\tilde{Z}^0}$\fi}
 \def\PSHz{\ifmmode\mathrm{\tilde{H}^0_j}
           \else$\mathrm{\tilde{H}^0_j}$\fi}
 \def\PSgxpm{\ifmmode\mathrm{\tilde{\chi}^{\pm_i}}
             \else$\mathrm{\tilde{\chi}^{\pm_i}}$\fi}
 \def\PSWpm{\ifmmode\mathrm{\tilde{W}^{\pm}}
            \else$\mathrm{\tilde{W}^{\pm}}$\fi}
 \def\PSHpm{\ifmmode\mathrm{\tilde{H}^{\pm_j}}
            \else$\mathrm{\tilde{H}^{\pm_j}}$\fi}
 \def\PSgn{\ifmmode\mathrm{\tilde{\nu}}
           \else$\mathrm{\tilde{\nu}}$\fi}
 \def\PSe{\ifmmode\mathrm{\tilde{e}}
          \else$\mathrm{\tilde{e}}$\fi}
 \def\PSgm{\ifmmode\mathrm{\tilde{\mu}}
           \else$\mathrm{\tilde{\mu}}$\fi}
 \def\PSgt{\ifmmode\mathrm{\tilde{\tau}}
           \else$\mathrm{\tilde{\tau}}$\fi}
 \def\PSq{\ifmmode\mathrm{\tilde{q}}
          \else$\mathrm{\tilde{q}}$\fi}
 \def\PSg{\ifmmode\mathrm{\tilde{g}}
          \else$\mathrm{\tilde{g}}$\fi}

\def\ks{$K^0_s$}
\def\ksp{$K^0_s p$}
\def\la{$\Lambda$}
\def\tp{\ifmmode\mathrm{\Theta^+}
          \else$\mathrm{\Theta^+}$\fi}
\def\t1540{\ifmmode{\mathrm \Theta(1540)^+}
           \else${\mathrm \Theta(1540)^+}$\fi}
\def\p1860{\ifmmode{\mathrm \Phi(1860)^{--}}
           \else${\mathrm \Phi(1860)^{--}}$\fi}
\def\xf{$x_F$}
\def\pt{$p_t$}

\def\pbar{\ifmmode\mathrm{\overline{p}}
         \else$\mathrm{\overline{p}}$\fi}
\def\pbarp{\ifmmode\mathrm{\overline{p}p}
         \else$\mathrm{\overline{p}p}$\fi}
\def\pbarn{\ifmmode\mathrm{\overline{p}n}
         \else$\mathrm{\overline{p}n}$\fi}

\def\Cascade{\ifmmode\mathrm{\Xi}
           \else$\mathrm{\Xi}$\fi}
\def\Cascadebar{\ifmmode\mathrm{\overline{\Xi}}
           \else$\mathrm{\overline{\Xi}}$\fi}

\def\Kp{\ifmmode\mathrm{K^+}
          \else$\mathrm{K^+}$\fi}
\def\Km{\ifmmode\mathrm{K^-}
          \else$\mathrm{K^-}$\fi}

\def\Panda{{\sc{Panda}}}
\def\HESR{{\sc{Hesr}}}
\def\HypHI{{\sc{HypHI}}}
\def\Agata{{\sc{Agata}}}

\newcommand{\goo}{\,\raisebox{-.5ex}{$\stackrel{>}{\scriptstyle\sim}$}\,}
\newcommand{\loo}{\,\raisebox{-.5ex}{$\stackrel{<}{\scriptstyle\sim}$}\,}
%--------------------------------------------------------------------------------
%\renewcommand{\floatpagefraction}{.99}% vorher: .5
\renewcommand{\textfraction}{.05} % vorher: .2
\renewcommand{\topfraction}{.95}     % vorher: .7
%\renewcommand{\bottomfraction}{.5}  % vorher: .3
% --------------------- end of definitions ------------------------------------

%---------------------------------------------------------------------------
\begin{frontmatter}



\title{Production of excited double hypernuclei via Fermi breakup of excited strange systems}


\author[Label1]{Alicia Sanchez Lorente}
\author[Label1,Label2]{Alexander S.~Botvina}
\author[Label1]{Josef~Pochodzalla}

\ead{pochodza@kph.uni-mainz.de}
\address[Label1]{Johannes Gutenberg-Universit{\"a}t Mainz, Institut f{\"u}r Kernphysik, D-55099 Germany}
\address[Label2]{Institute for Nuclear Research, Russian Academy of Sciences,
117312 Moscow, Russia}

\begin{abstract}
Precise spectroscopy of multi-strange hypernuclei provides a unique
chance to explore the hyperon-hyperon interaction. In the present
work we explore the production of excited states in double
hypernuclei following the micro-canonical break-up of an initially excited
double hypernucleus which is created  by the absorption and conversion of a stopped
$\Xi^{-}$ hyperon. Rather independent on the spectrum of possible
excited states in the produced double hypernuclei the formation of
excited states dominates in our model. For different initial
target nuclei which absorb the  $\Xi^-$,  different double hypernuclei
nuclei dominate. Thus the ability to assign the various observable
$\gamma$-transitions in a unique way to a specific double
hypernuclei by exploring various light targets as proposed by
the {\Panda} collaboration seems possible. We also confront our predictions with the correlated pion spectra measured by the E906 collaboration.
\end{abstract}
\end{frontmatter}

While the nucleon-nucleon scattering was extensively studied since
the 50's, direct experimental investigations for the hyperon-hyperon
interactions are still very sparse. Because of their short
lifetimes, hyperon targets are not available and low momentum
hyperons are very difficult to produce. Since direct scattering
experiments between two hyperons are impractical, the precise
spectroscopy of multi-strange hypernuclei provides a unique chance
to explore the hyperon-hyperon interaction. Indeed the significant
progress in nuclear structure calculations
\cite{Wir02,Pie04,Pie05,chi00,chi01} nurtures the hope that detailed
information on excitation spectra of double hypernuclei and their structure
will provide unique information on the hyperon-hyperon interactions.

The simultaneous production and implementation of two $\Lambda$
particles into a nucleus is intricate. {\color{black}There may be a
chance to produce multi-strange hypernuclei in hadron and heavy ion induced
collisions \cite{Ker73,Bot07,Gai09}, as it was
impressively illustrated by the STAR collaboration recently \cite{STAR}.
However, high resolution spectroscopy of excited states will not not feasible.
While in central collisions statistical approaches \cite{And10} may be applicable, the production of hypernuclei in peripheral heavy ion collisions clearly requires a detailed consideration of the collision dynamics to make definite predictions of the production probability \cite{Gai09}.}
To produce double hypernuclei in a more `controlled' way the conversion
of a captured $\Cascade^-$ and a proton into two $\Lambda$ particles
can be used. This process releases -- ignoring binding energy
effects -- only 28\,MeV. For light nuclei there exists therefore a
significant probability of the order of a few percent that both
$\Lambda$ hyperons are trapped in the same excited nuclear
fragment~\cite{Yam94,Yam97,Yam97b,Hir97,Hir99,Aok09}.

Unfortunately $\Cascade^-$ hyperons produced in reactions with
stable hadron beams have usually rather high momenta. Therefore,
direct capture of the $\Cascade^-$ in the nucleus is rather
unlikely. Even in case of the {\color{black}$(\Km, \Kp)$} double strangeness
exchange reaction, $\Cascade^-$ hyperons are produced with typical
momenta of 500 {\mevc1} at beam momenta around 1.8 GeV/c
\cite{Yam97b,Nar98,Ohn01}. The advantage of this production process is that
the outgoing {\color{black} $\Kp$} can be used as a tag for the reaction. A drawback
is the low kaon beam intensity and hence the need for thick primary
targets. Furthermore, as a consequence of the large momentum
transfer, the probability to form bound $\Xi^-$ states directly is
rather small on the level of 1$\%$ \cite{Ike94,Yam94b} and the
production of quasi-free  $\Xi^-$ dominates. Still the $\Xi^-$
hyperons in the quasi-free region may be absorbed into the target
nucleus via a rescattering process on a nucleon which itself is
knocked out of the primary nucleus. This two-step process is
predicted to exceed the direct capture by more than a factor of 6
\cite{Yam97} (see also below).

%-------------------------------------Table 1 ----------------------
\begin{table}[h]
\caption{Ground and exited states of double hypernuclei used in the
present calculation together with the relevant references.}
\begin{tabular}{|c|c|c|c|c|}
\hline
Nucleus & mass g.s  & exc. energy & $J^{P}$ & Ref./note\\
   ~    & (\Mevc2)  & (MeV)       &  ~  & ~\\ \hline
$^{4}_{\Lambda\Lambda}$H & 4107.47 &  0.0  & $1^{+}$ & \cite{Ahn01,Ran07,Nem05} \\
& & & & \cite{Fil02,Nem03,Sho05} \\
\hline
$^{5}_{\Lambda\Lambda}$H & 5037.54 &  0.0  & $\frac{1}{2}^{+}$  & \cite{Myi03,Fil03,Lan04}\\
& & & & \cite{Sho04,Nem05}\\
\hline
$^{5}_{\Lambda\Lambda}$He & 5036.98 &  0.0  & $\frac{1}{2}^{+}$  & \cite{Myi03,Fil03,Lan04}\\
& & & & \cite{Sho04,Nem05}\\
\hline
$^{6}_{\Lambda\Lambda}$He & 5951.37 &  0.0  & $0^{+}$ & \cite{Tak01,Hiy02} \\
\hline
$^{7}_{\Lambda\Lambda}$He & 6889.82 & 0.0 & $\frac{3}{2}^{-}$ & \cite{Hiy02}\\
\hline
$^{8}_{\Lambda\Lambda}$He & 7825.18 & 0.0 & $0^+$ &  ${\Delta}B_{\Lambda\Lambda}$= \\
& &  1.80  &$ 2^{+} $ & 1\,MeV\\
\hline
$^{9}_{\Lambda\Lambda}$He & 8761.60 & 0.0 & $\frac{3}{2}^{-}$ &  ${\Delta}B_{\Lambda\Lambda}$= \\
& &  2.92  &$ \frac{5}{2}^{-} $ &  1\,MeV\\
\hline
$^{7}_{\Lambda\Lambda}$Li  & 6889.35 & 0.0 & $\frac{3}{2}^{-}$ &  \cite{Hiy02}\\
\hline
$^{8}_{\Lambda\Lambda}$Li  & 7821.27 &  0.0&  $1^{+}$ &  \cite{Hiy02} \\
& &  1.36  &$ 3^{+} $ &  \cite{Hiy02}\\
& &  5.63  &$ 2^{+} $ &  \cite{Hiy02}\\
\hline
$^{9}_{\Lambda\Lambda}$Li  & 8748.85 &  0.0     & $\frac{3}{2}^{-}$ &  \cite{Hiy02}\\
& & 0.73  &$\frac{1}{2}^{-}$ &  \cite{Hiy02}\\
& & 4.55  &$\frac{7}{2}^{-}$ &  \cite{Hiy02}\\
& & 5.96  &$\frac{5}{2}^{-}$ &  \cite{Hiy02}\\
\hline
$^{9}_{\Lambda\Lambda}$Be  & 8751.04  &  0.0     &$\frac{3}{2}^{-}$ &  \cite{Hiy02}\\
& & 0.71  &$\frac{1}{2}^{-}$ &  \cite{Hiy02}\\
& & 4.54  &$\frac{7}{2}^{-}$ &  \cite{Hiy02}\\
& & 5.92  &$\frac{5}{2}^{-}$ &  \cite{Hiy02}\\
\hline
$^{10}_{\Lambda\Lambda}$Be  & 9671.08 &  0.0     &$0^{+}$ &  \cite{Hiy02}\\
& & 2.86  &$2^{+}$ &  \cite{Hiy02}\\
\hline

$^{11}_{\Lambda\Lambda}$Be  & 10603.35  &  0.0     &$\frac{3}{2}^-$ & \cite{Yam97}\\
& & 1.684  &${\frac{1}{2}}^+$ & 9 excited\\
%& & 2.429  &${\frac{5}{2}}^-$ & states\\
& & ...  & ... & states\\
& & 7.94  &${\frac{5}{2}}^-$ & \\

\hline
$^{11}_{\Lambda\Lambda}$B  & 10603.71 &  0.0     &$\frac{3}{2}^{-}$ &  \cite{Yam97}\\
& & 2.361 &$\frac{5}{2}^{-}$ & \\
& & 2.75  &$\frac{1}{2}^{-}$ & \\
& & 2.788  &$\frac{5}{2}^{+}$& \\

\hline
$^{12}_{\Lambda\Lambda}$Be  & 11534.05 &  0.0     &$0^{+}$ &\cite{Yam97}\\
& & 3.36  &$2^+$ &  7 excited\\
%& & 5.958  &$2^+$ & states\\
& & ...  & ... & states\\
& & 7.542  &$2^+$ & \\

\hline
$^{12}_{\Lambda\Lambda}$B  & 11532.52 &  0.0  & $3^{+}$    & \cite{Yam97}\\
& & 0.71835  & $1^+$  & 24 excited  \\
%& & 1.74015  & $0^+$  & states  \\
& & ...  & ... & states\\
& & 8.894  & $2^+$ & \\
\hline

$^{13}_{\Lambda\Lambda}$B  & 12456.59 &  0.0 &$\frac{3}{2}^{-}$
&\cite{Yam97}
\\\hline
\end{tabular}
\label{tab01}
\end{table}
%---------------------------------------Table 1 -----------------

On the other hand most ($\sim 80\% $) $\Xi^-$ hyperons escape from
the primary target nucleus in {\color{black} $(\Km, \Kp)$} reactions. However, in a
second step, these $\Cascade^-$ hyperons can be slowed down in a
dense, solid material (e.g.\ a nuclear emulsion) and form
$\Cascade^-$ atoms~\cite{Bat99}. After an atomic cascade, the
$\Xi$-hyperon is eventually captured by a secondary target nucleus
and converted via the $\Xi^-p\rightarrow \Lambda\Lambda$ reaction
into two $\Lambda$ hyperons. In a similar two-step process
relatively low momentum $\Cascade^-$ can also be produced in $\pbarp
\rightarrow \Cascade^- \Cascadebar^+$ or $\pbarn \rightarrow
\Cascade^- \Cascadebar^\circ$ reactions if this reactions happens in
a complex nucleus where the produced $\Xi^-$ can re-scatter
\cite{Poc04,PandaTPR}. The advantage as compared to the kaon induced
reaction is that antiprotons are stable and can be
retained in a storage ring. This allows a rather high luminosity
even with very thin primary targets.

Because of the two-step mechanism, spectroscopic studies, based on
two-body kinematics like in single hypernucleus production, cannot
be performed. Spectroscopic information on double hypernuclei can
therefore only be obtained via their decay products. The kinetic
energies of weak decay products are sensitive to the binding
energies of the two $\Lambda$ hyperons.
While the double pionic decay of light double hypernuclei can be used
as an effective experimental filter to reduce the background
\cite{PandaPhysicsbook}
the unique identification of hypernuclei groundstates only via their pionic decay is usually hampered by the limited resolution (see e.g. ref.~\cite{Ran07} and discussion below). Instead, $\gamma$-rays emitted via the sequential decay of excited double hypernuclei may provide precise
information on the level structure.

% fig.1 --------------------------------------------------------
\begin{figure}[t]
\includegraphics[width=1.0\linewidth]{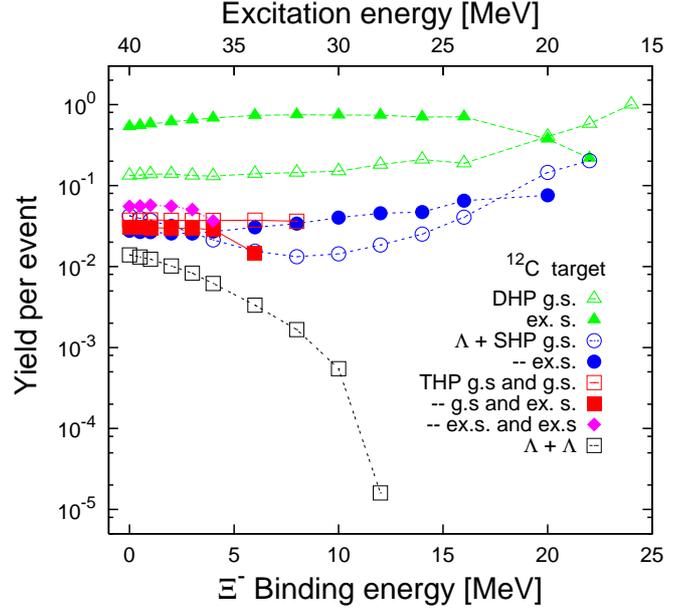}
\caption{Predicted production probability of ground (g.s.) and excited states
(ex.s.) in one single (SHP), twin (THP) and double hypernuclei (DHP)
after the capture of a $\Xi^-$ in a $^{12}$C nucleus and its
conversion into two $\Lambda$ hyperons . The lower and upper scale shows the binding energy of the captured $\Xi^-$ and the excitation energy of the initial $^{13}_{\Lambda\Lambda}$B nucleus, respectively.}
  \label{fig:sanchez01}

\end{figure}
% --------------------------------------------------------------

The \Panda\ experiment \cite{PandaTPR} which is planned at the
international Facility for Antiproton and Ion Research {\sc FAIR} in
Darmstadt aims at the high resolution $\gamma$-ray spectroscopy of
double hypernuclei \cite{Poc04}. {\color{black} In this work we study the important question to what extend
particle stable excited states of double hypernuclei are produced and how different secondary target
nuclei by help to assign observed $\gamma$-transitions.}
In line with the compound nucleus
model of Sano and co-workers \cite{San92,Yam94} we study in the
present work the production of excited states in double hypernuclei
following the break-up of an excited double hypernucleus after the
absorption and conversion of a stopped $\Xi ^{-}$.

For light nuclei even a relatively small excitation energy may  be comparable with
their binding energy. We therefore consider the explosive decay of
the excited nucleus into several smaller clusters as the principal
mechanism of de-excitation. {\color{black} Not included in the present approach are neither
details of the atomic capture process nor of a possible dynamical, non-equilibrium stage
prior to the $\Xi^-$ conversion into two $\Lambda$ hyperons.
Unlike in case of a direct $\Xi^-$ capture in $(\Km, \Kp)$ or other hadron induced reactions where a pre-equilibrium stage will require special attention (see e.g. Ref.~\cite{Gai09}), pre-equilibrium processes are probably less important after an atomic capture.}\footnote{\label{foot:1}
With respect to the number of stopped, $\Xi^-$ hyperons, pre-equilibrium processes will
decrease the yield of double hypernuclei relative to
the yield for single hypernuclei (see e.g. \cite{Aok09}). Indeed in the
simulations for the planned \Panda\ experiment \cite{PandaPhysicsbook}
a joint capture$\times$conversion probability of 5$\%$ was assumed
to mimic this pre-equilibrium stage. Taking this factor into account
the probability for the various channels is compatible with present
scarce experimental information \cite{Hir97,Hir99,Aok09}.}

To describe this break-up process we have developed a model which is
similar to the famous Fermi model for particle production in nuclear
reactions \cite{Fermi}. We assume that the nucleus with mass numbers
$A_0$, charge $Z_0$, and the number of $\Lambda$ hyperons $H_0$
(here $H_0$=2) break-ups simultaneously into cold and slightly
excited fragments, which have a lifetime longer than the decay time,
estimated as an order of 100-300 fm/c. This break-up takes place in
some freeze-out volume $V$, where the produced fragments moves in
the phase space determined by the free volume $V_{f}$. This free
volume is smaller than the freeze-out volume, at least, by the
proper volume of the fragments. We use the `excluded volume'
approximation for this parametrization $V=V_{f}+V_0$, where
$V_0=\frac{4}{3}\pi r_0^3 A_0$ denotes the initial volume of the
nucleus  with $r_0= 1.3$ fm and $V_{f}=\kappa \cdot V_0$. In the
following we assume for the parameter $\kappa$ a value of 2 which is
consistent with many description of fragmentation data of normal
nuclei \cite{SMM}.

In the case of production of conventional nuclear fragments in a
break-up channel, we adopt their experimental masses in ground
states, and take into account their excited states, which are stable
respective to emission of nucleons \cite{Ajzenberg}. For hypernuclei
with single $\Lambda$ particle, we use the experimental masses and
excited states, which were collected in various reviews
\cite{Bando,Has06} and also some theoretical predictions\cite{Hiy09}.
One has to keep in mind that particularly for heavy hypernuclei
additional excited states may be missing in the present decay channels.

For double hypernuclei the experimental information is restricted to
a few cases only \cite{Dan63a,Pro66,Aok91a,Ahn01,Tak01,Aok09}. Except for
the $^6_{\Lambda\Lambda}$He nucleus reported in Ref. \cite{Tak01}
the interpretation of the observed events is however not unique
\cite{Dan63b,Dal89,Dov91,Hiy02,Ran07}. Furthermore no direct
experimental information on possible excited states is at hand (see
e.g. discussion in Ref. \cite{Hiy02}). Therefore, theoretical
predictions of bound and exited states of double hypernuclei
predicted by Hiyama and co-workers \cite{Hiy02} were used in the
present model calculation for nuclei with mass number 6$\leq
A_0\leq$10. The masses of $^8_{\Lambda\Lambda}$He and $^9_{\Lambda\Lambda}$He
were calculated from the known single hypernuclear masses and assuming ${\Delta}B_{\Lambda\Lambda}$=1\,MeV.
Here ${\Delta}B_{\Lambda\Lambda}$ is defined in terms of the involved nuclear masses
\begin{equation}
{\Delta}B_{\Lambda\Lambda}=2M(^{A-1}_{\Lambda}Z)-M(^{A-2}Z)-M(^A_{\Lambda\Lambda}Z)
\end{equation}
and which contains information on the binding between the two $\Lambda$ hyperons.
For the mirror nuclei $^5_{\Lambda\Lambda}$H and
$^5_{\Lambda\Lambda}$He there seems to be a consensus that these
nuclei are indeed bound \cite{Myi03,Fil03,Lan04,Sho04,Nem05}. In
view if the theoretical uncertainties, we again assumed in our
calculations a value of ${\Delta}B_{\Lambda\Lambda}$=1\,MeV for both
nuclei. In case of $^4_{\Lambda\Lambda}$H the experimental situation
is ambiguous \cite{Ahn01,Ran07} and also various model calculations
predict an unbound \cite{Fil02} or only slightly bound nucleus
\cite{Nem03,Sho05,Nem05}. We checked however that our results remain
unchanged for the production of heavier nuclei if we disregard this
slightly bound $^4_{\Lambda\Lambda}$H nucleus.

% fig.2 --------------------------------------------------------

\begin{figure}[t]
\includegraphics[width=0.9\linewidth]{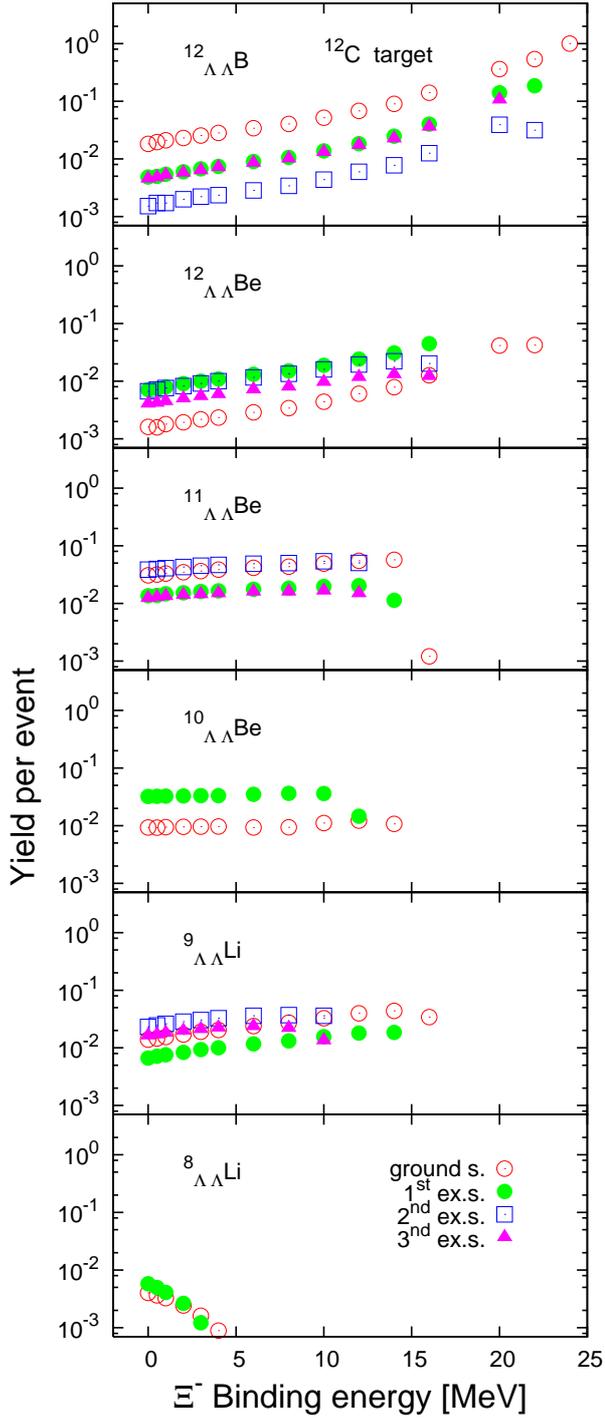}
\caption{Individual production probability of up to four lowest lying,
particle stable states in double hypernuclei after the capture of a
$\Xi^-$ by a $^{12}$C nucleus.} \label{fig:sanchez02}
\end{figure}

% --------------------------------------------------------------
For nuclei with mass number A$\geq$11 the masses of the ground
states were taken from ref.~\cite{Yam97}. These masses correspond to ${\Delta}B_{\Lambda\Lambda}$ values between about 2.0 and 4.6\,MeV.
Also for these heavy nuclei several particle stable excited states are expected
\cite{Yam97}. The calculations of Hiyama and co-workers \cite{Hiy02}
signal that in the mass range relevant for this work the level
structure of particle stable double hypernuclei resembles the level
scheme of the corresponding core nucleus. In order to explore the
role of possible additional states the set of double hypernuclei
with A$\geq$11 was extended by the known excited states of the
corresponding core nucleus. Only states below the lowest particle
decay threshold were considered. In the present calculations $1s_{\Lambda}1p_{\Lambda}$ states have been
ignored though these states may possibly contribute at high excitation energies in heavy nuclei \cite{Dov94}. In case of
$^{11}_{\Lambda\Lambda}$Be, $^{11}_{\Lambda\Lambda}$B,
$^{12}_{\Lambda\Lambda}$Be and $^{12}_{\Lambda\Lambda}$B this recipe
resulted in 9, 3, 7 and 24 excited states, respectively
(Tab.~\ref{tab01}).
Within the model systematic errors may arise due to missing states in single and double hyper nuclei. In order to check the sensitivity of our conclusions to the modifications of
the allowed decay products, we have repeated the calculations by only considering the ground and at most two particle stable, excited states in double hyperfragments.
The relative yield of some individual excited states was modified up to a factor 2$^{\pm1}$. However, the final conclusions of our paper do not not change.

In the model we consider all possible break-up channels, which
satisfy the mass number, hyperon number (i.e. strangeness), charge,
energy and momentum conservations, and take into account the
competition between these channels. The model assumes that the
probability of each break-up channel $ch$ is proportional to the
occupied phase space \cite{SMM,Botvina87,preprint90}. The
statistical weight of the channel containing $n$  particles  with
masses $m_{i}$ ($i=1,\cdots,n$) can be calculated in microcanonical
approximation:
\begin{eqnarray}
\label{eq:Fer} W_{ch}^{mic}\propto
\frac{S}{G}\left(\frac{V_{f}}{(2\pi\hbar)^{3}}\right)^{n-1}
\left(\frac{\prod_{i=1}^{n}m_{i}}{m_{0}}\right)^{3/2}\\
~\cdot \frac{(2\pi)^
{\frac{3}{2}(n-1)}}{\Gamma(\frac{3}{2}(n-1))}\cdot
\left(E_{kin}-U_{ch}^{C}\right)^{\frac{3}{2}n-\frac{5}{2}},
\end{eqnarray}
where $m_{0}=\sum_{i=1}^{n}m_{i}$ is the summed mass of the
particles, $S=\prod_{i=1}^{n}(2s_{i}+1)$ is the spin degeneracy
factor ($s_{i}$ is the $i$-th particle spin),
$G=\prod_{j=1}^{k}n_{j}!$ is the particle identity factor ($n_{j}$
is the number of particles of kind $j$). $E_{kin}$ is the total
kinetic energy of  particles at infinity which is related to the
available energy via
\begin{equation}
\label{eq:Ek}
E_{kin}=(M(\Xi^{-})+M_{target})c^2-B_{\Xi}-\sum_{i=1}^{n}m_{i}c^2.
\end{equation}
Here B$_{\Xi}$ is the binding energy of the converted $\Xi^-$ and
$E_{max}=(M(\Xi^{-})+M_{target})c^2$ represents the maximum
available excitation energy ignoring the binding of the $\Xi^-$,
$U_{ch}^{C}$ is the Coulomb interaction energy between fragments
given in the Wigner-Seitz approximation \cite{SMM}:
\begin{equation}
\label{eq:UjC}
U_{ch}^{C}=\frac{3}{5}\frac{e^{2}}{r_0}(V/V_0)^{-1/3}
(\frac{Z_{0}^{2}}{A_0^{1/3}}-\sum_{i=1}^{n}\frac{{Z_i}^{2}}{{A_i}^{1/3}}),
\end{equation}
where $A_i$, $Z_i$ are mass numbers and charges of produced
particles. We calculate masses of fragments in excited states by
adding the corresponding excitation energy to their ground state
masses.

Unfortunately, still very little is established experimentally on
the interaction of $\Xi$ hyperons with nuclei. Various data suggest
a nuclear potential well depth around 20\,MeV (see e.g.
\cite{Dov83,Fri07}). Calculations of light $\Xi$ atoms \cite{Bat99}
predict that the conversion of the captured $\Xi^-$ from excited
states with correspondingly small binding energies dominates. In a
nuclear emulsion experiment a $\Xi^-$ capture at rest with two
single hyperfragments has been observed \cite{Aok95} which was
interpreted as $\Xi^-$ + $^{12}$C $\rightarrow ^4_{\Lambda}$H +
$^9_{\Lambda}$Be reaction. The deduced binding energy of the $\Xi^-$
varied between 0.62\,MeV and 3.70\,MeV, depending whether only one
out of the two hyperfragments or both fragments were produced in an
excited particle stable state. Model calculations
\cite{Ike94,Mil94,Dov94,Yam06} suggest that the width for the $\Xi ^{-}+p
\rightarrow \Lambda+\Lambda$ conversion is around 1 MeV, i.e. the
conversion is rather fast and takes of the order of 200 fm/c. In
order to take into account the uncertainties of the excitation
energy of the converted $\Xi^-$-states, the calculations were
performed for a range of energies $0 \leq B_{\Xi} \leq E_{max}$,
constructing in this way the excitation functions for the production
of hypernuclei.

The Fermi break-up events were generated by comparing probabilities
of all possible channels with Monte--Carlo method. For example a
total of 993 channels were included in case of the absorption of a
$\Xi ^{-}$ on $^{13}$C nucleus. The Coulomb expansion stage was not considered
explicitly for such light systems. The momentum distributions of the
final break-up products were obtained by the random generation over
the whole accessible phase space, determined by the total kinetic
energy (\ref{eq:Ek}), taking into account exact energy and momentum
conservation laws. For this purpose we applied a very effective
algorithm proposed by G.I.~Kopylov \cite{Kopylov}.

Previously, this model was applied rather successfully for the
description of the break-up of conventional light nuclei in nuclear
reactions initiated by protons, pions, antiprotons, and ions
\cite{SMM,preprint90,Tashkent,Sudov}. In comparison with experimental yields
of different fragments the maximum deviation was about 20--50\%.
We believe that this precision is sufficient for the
present analysis of hypernucleus decays, since the main uncertainly
is in unknown masses and energy levels of produced hyperfragments.

%----------- fig.3 ---------------------------------------------
\begin{figure}[t]
\includegraphics[width=1.0\linewidth]{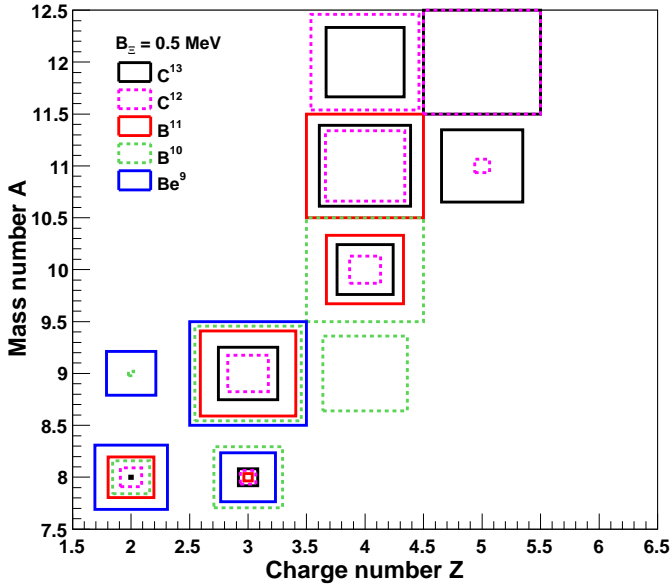}
\caption{Relative population of the summed first + second {\em excited}
states in all produced double hypernuclei for various
$\Xi$-absorbing stable target nuclei $^9$Be, $^{10}$B, $^{11}$B,
$^{12}$C and $^{13}$C. The area of the squares is proportional to the
probability. For each secondary target the area for the most likely excited nucleus
is normalized to 1.} \label{fig:sanchez03}
\end{figure}
% --------------------------------------------------------------

\begin{table*}[hbt]
\caption{Total production probability of particle stable twin and
double hypernuclei after the capture of a $\Xi^-$ by a $^9$Be target
and the conversion into an excited $^{10}_{\Lambda\Lambda}$Li$^*$
hypernucleus (third and forth \cite{Yam97b} column). The four last
columns are the results assuming the production of excited
$^8_{\Lambda\Lambda}$He$^*$ and $^8_{\Lambda\Lambda}$H$^*$ nuclei
after a knock-out process with an excitation energy of 33\,MeV
\cite{Yam97}. Here columns 5 and 6 are results of the present work, the
last two columns are again from Ref.~\cite{Yam97b}. A -- indicates
that this particular channel cannot be reached or was not
considered.}

\begin{tabular}{|c|cc||c|c||c|c||c|c|}
\hline decay channel & \multicolumn{2}{c|}{$\pi$ pair momenta} & \multicolumn{6}{c|}{decaying system and probability} \\
 ~& \multicolumn{2}{c|}{(MeV/c)} & $^{10}_{\Lambda\Lambda}$Li$^*$ & $^{10}_{\Lambda\Lambda}$Li$^*$ \cite{Yam97b}&   $^{8}_{\Lambda\Lambda}$He$^*$  & $^{8}_{\Lambda\Lambda}$H$^*$
 & $^{8}_{\Lambda\Lambda}$He$^*$ \cite{Yam97b}& $^{8}_{\Lambda\Lambda}$H$^*$ \cite{Yam97b}
\\ \hline
$^{3}_{\Lambda}$H+$^3_{\Lambda}$H              & 114& 114  & --       & 0      & 0      & --    & 0     & --\\
$^{3}_{\Lambda}$H+$^4_{\Lambda}$H$_{gs}$       & 114& 133  & --       & 0      & 0.008  & --    & 0.018 & --\\
$^{3}_{\Lambda}$H+$^4_{\Lambda}$H$_{1.05}$     & 114& 134  & --       & --     & 0.014  & --    & --    & --\\
$^{3}_{\Lambda}$H+$^5_{\Lambda}$He             & 114& 99   & 0.0001   & 0.011  &--      & --    & --    & --\\
$^{3}_{\Lambda}$H+$^6_{\Lambda}$He             & 114& 108  & 0.004    & 0.012  &--      & --    & --    & --\\
$^{3}_{\Lambda}$H+$^7_{\Lambda}$He$_{gs}$      & 114& 115  & 0.026    &  0.018 &-- & -- & --    & --\\
$^{3}_{\Lambda}$H+$^7_{\Lambda}$He$_{1.66}$    & 114& 118  & 0.046    &  0.018 &-- & -- & --    & --\\
$^{3}_{\Lambda}$H+$^7_{\Lambda}$He$_{1.74}$    & 114& 118  & 0.068    &  0.018 &-- & -- & --    & --\\
$^{4}_{\Lambda}$H$_{gs}$+$^4_{\Lambda}$H$_{gs}$    & 133& 133  & --       & 0.005  & 0.017  & --    & 0.055 & --\\
$^{4}_{\Lambda}$H$_{gs}$+$^4_{\Lambda}$H$_{1.05}$  & 133& 134  & --       & --     & 0.096  & --    & --    & --\\
$^{4}_{\Lambda}$H$_{1.05}$+$^4_{\Lambda}$H$_{1.05}$& 134& 134  & --       & --     & 0.137  & --    & --    & --\\
$^{4}_{\Lambda}$H$_{gs}$+$^5_{\Lambda}$He          & 133& 99   & 0.022    & 0.045  &--      & --    & --    & --\\
$^{4}_{\Lambda}$H$_{1.05}$+$^5_{\Lambda}$He        & 134& 99   & 0.055  & --  &--      & --    & --    & --\\
$^{4}_{\Lambda}$H$_{gs}$+$^6_{\Lambda}$He          & 133& 108  & 0.031    & 0.049  &--      & --    & --    & --\\
$^{4}_{\Lambda}$H$_{1.05}$+$^6_{\Lambda}$He        & 134& 108  & 0.088    & --  &--      & --    & --    & --\\ \hline
$^{4}_{\Lambda\Lambda}$H                           & 117& 98   & 0.0006  & 0.026  & 0.003  & 0.0002& 0.026 & 0 \\
$^{5}_{\Lambda\Lambda}$H                           & 134& 99   & 0.007   & 0.139  & 0.069  & 0.635 & 0.108 & 0.877 \\
$^{5}_{\Lambda\Lambda}$He                    & -- &--    & --      & 0.0  & 0      & --    & 0.009 & -- \\
$^{6}_{\Lambda\Lambda}$He                 & 100&99 \cite{6HeL}& 0.028  & 0.147  & 0.051  & --    & 0.128 & -- \\
$^{7}_{\Lambda\Lambda}$He                    & 109& 108  & 0.117   & 0.133  & 0.116  & --    & 0.157 & -- \\
$^{8}_{\Lambda\Lambda}$He$_{gs}$             & 116& 124  & 0.022   & small  & --     & --    & --    & -- \\
$^{8}_{\Lambda\Lambda}$He$_{ex}$             & 119& 124  & 0.096   & --     & --     & --    & --    & -- \\
$^{9}_{\Lambda\Lambda}$He$_{gs}$             & 117& 121  & 0.021   & 0.025  & --     & --    & --    & -- \\
$^{9}_{\Lambda\Lambda}$He$_{ex}$             & 122& 121  & 0.027   & --     & --     & --    & --    & -- \\
$^{7}_{\Lambda\Lambda}$Li                    & 101& 96   & 0.0001   & 0.008  & --     & --    & --    & -- \\
$^{8}_{\Lambda\Lambda}$Li$_{gs}$             & 109& 97   & 0.012    & 0.028  & --     & --    & --    & -- \\
$^{8}_{\Lambda\Lambda}$Li$_{ex}$             & 111-117&97& 0.028    & --     & --     & --    & --    & -- \\
$^{9}_{\Lambda\Lambda}$Li$_{gs}$             & 123& 97   & 0.028    & 0.026  & --     & --    & --    & --\\
$^{9}_{\Lambda\Lambda}$Li$_{ex}$             & 124-131&97& 0.098& -- & --     & -- & -- & -- \\\hline
\end{tabular}
\label{tab02}
\end{table*}

Fig.~\ref{fig:sanchez01} shows as an example the production of
ground (g.s.) and excited (ex.s.) states of single + one free $\Lambda$ (SHP), twin (THP)
and double (DHP) hypernuclei in case of a $^{12}$C target as a
function of the assumed $\Xi^-$ binding energy.
With increasing $\Xi^-$ binding energy the excitation energy of the excited primary $^{13}_{\Lambda\Lambda}$B$^*$ nucleus decreases from left to right from about 40\,MeV to 15\,MeV. For all excitation energies above 20\,MeV the production of {\em
excited} double hypernuclei dominates (green triangles). This can be
traced back to the opening of several thresholds for various excited
double hypernuclei already at moderate excitation energies (see e.g.
Fig.~1 in Ref~\cite{Yam97}). Only for small binding energies and
hence large excitation energies the production of single and twin
hypernuclei is significant ($\sim$10$\%$). The
non-monotonic behaviour for single hypernucleus + one free $\Lambda$
production reflects the fact that the various lowest thresholds are
relative high and widely separated, e.g. $^{12}_{\Lambda}B+\Lambda$
at B$_{\Xi}$=23.9\,MeV followed by $^{11}_{\Lambda}B+n+\Lambda$ at
11.3\,MeV. Twin-hypernuclei are only
produced for $\Xi^-$ binding energies below the threshold for
$^8_{\Lambda}Li + ^5_{\Lambda}He$ with B$_{\Xi}$=13.6\,MeV. As
discussed above the frequent observation of twin-hypernuclei
\cite{Wil59,Ste63,Bec68,Aok93,Aok95,Ich01} signals a conversion from
a $\Xi$ state with only moderate binding energy. In this range of
B$_{\Xi}$ the production probability of double hypernuclei is
comparable to previous estimates within a canonical statistical
model \cite{Yam97,Yam97b}.

%----------- fig.4 ---------------------------------------------
%\begin{figure*}[t]
%\includegraphics[width=0.9\linewidth]{Sanchez_Fig4f.eps}
%\caption{Relative production probability of double (top part) and twin hypernuclei (lower part) for %a primary $^{10}_{\Lambda\Lambda}$Li nucleus as a function of its excitation energy.}
%\label{fig:sanchez04}
%\end{figure*}
% --------------------------------------------------------------

Fig.~\ref{fig:sanchez02} shows the population of the four lowest
lying, particle stable states in double hypernuclei. The
$^{12}_{\Lambda\Lambda}$Be+p and $^{12}_{\Lambda\Lambda}$B+n have
the lowest thresholds \cite{Yam97} and dominate at large binding
energies and hence small excitation energies. At smaller and more
realistic binding energies the lighter double hypernuclei
$^{11}_{\Lambda\Lambda}$Be, $^{10}_{\Lambda\Lambda}$Be and
$^{9}_{\Lambda\Lambda}$Li take over. Due to the higher spin
degeneracy factor the population of excited states can exceed those
of ground states significantly at low binding energies. The dominant
population of excited states in e.g. $^{10}_{\Lambda\Lambda}$Be is
consistent with the conjecture of Hiyama and co-workers \cite{Hiy02}
that the Demachi-Yanagi event \cite{Demachi1,Demachi2} can be
interpreted most probably as the observation of the 2$^+$ excited
state in $^{10}_{\Lambda\Lambda}$Be. These calculations also indicate
that by studying relative probabilities of excited states one
may obtain information on the binding energy of the
captured $\Xi^-$ hyperons.

The summed population of the first and second particle stable excited
states in all produced double hypernuclei are shown in
Fig.~\ref{fig:sanchez03} for various $\Xi$-absorbing stable target
nuclei $^9$Be, $^{10}$B, $^{11}$B, $^{12}$C and $^{13}$C.
The probabilities are proportional to the area of the squares. For each target
the yield for the most probable hypernucleus has been normalized to 1.
For all five target nuclei the same $\Xi^-$ binding energy of 0.5\,MeV was
adopted. Although this choice is not crucial for the main trend
(see Fig.~\ref{fig:sanchez01}), the population of
specific excited levels may depend somewhat on the adopted binding energy.
In Fig.~\ref{fig:sanchez03} different double hypernuclei dominate for each target.
Thus combining the information of Fig.~\ref{fig:sanchez03} with the measurement of
two pion momenta from the subsequent weak decays a unique assignment of various newly observed
$\gamma$-transitions to specific double hypernuclei seems possible as intended by the {\Panda} collaboration \cite{Poc04,PandaPhysicsbook}.

In 2001 the BNL experiment E906 reported the observation of the
$^4_{\Lambda\Lambda}$H hypernucleus by measuring the sequential
pionic decays after a (K$^-$,K$^+$) reaction deposited two units of
strangeness in a $^9$Be target \cite{Ahn01}. Two structures
in the correlated $\pi^-$ momenta at (133,114)\mevc1 and at (114,104)\mevc1
were observed. The first structure was interpreted as the production of
$^3_{\Lambda}$H+$^4_{\Lambda}$H twins while the bump at (114,104)\mevc1 was attributed to pionic decays of the double hypernucleus $^4_{\Lambda\Lambda}$H. However, as it was pointed out by Kumagai-Fuse and Okabe also twin $\Lambda$-hypernuclear decays of $^3_{\Lambda}$H and $^6_{\Lambda}$He are a possible candidate to form this peak if
excited resonance states of $^6$Li are considered \cite{Kum02}. More
recently Randeniya and Hungerford showed that the published E906 data can be reproduced without the inclusion of $^4_{\Lambda\Lambda}$H decay and that it is
more likely that the decay of $^7_{\Lambda\Lambda}$He was observed
in the E906 experiment \cite{Ran07}. In their analysis this double
hypernucleus was accompanied by a background of coincident decays of single hypernuclei pairs $^3_{\Lambda}$H+$^4_{\Lambda}$H, $^3_{\Lambda}$H+$^3_{\Lambda}$H, and
$^4_{\Lambda}$H+$^4_{\Lambda}$H, respectively. Here the first twin pair
dominated and the production ratio $^{7}_{\Lambda\Lambda}$He to
coincident $^3_{\Lambda}$H+$^4_{\Lambda}$H pairs was estimated to be
in the range of 7.7 to 12.

Tab.~\ref{tab02} summarizes the predicted relative probabilities for
the  production of particle stable twin and double hypernuclei in the
E906 experiment. In the second column we list for orientation the
associated pion momenta, assuming a
$\Lambda\Lambda$ bond energy of ${\Delta}B_{\Lambda\Lambda}$=1\,MeV and subsequent pionic two-body decays from the excited or groundstate to the corresponding groundstate.
For orientation an increase of ${\Delta}B_{\Lambda\Lambda}$ to e.g. 4\,MeV decreases the first momenta given in the second column for $\Lambda\Lambda$ hypernuclei by about 5{\mevc1}.
Note that lower pion momenta may also arise from multibody decays or the decay into excited intermediate nuclei. Excited states which will {\color{black} be} deexcited prior to the weak decay will of course also result in lower pion momenta.
Comparing to experimental data one also has to  keep in mind that starting from pion pairs with rather similar or equal pion momenta, a finite momentum resolution will due to the sorting of the momenta result in a more asymmetric peak structure. In case of the $^{7}_{\Lambda\Lambda}He$ decays a momentum resolution like in the E906 experiment of  $\sigma_{rms}$=4\mevc1 will shift the peak from (109,108)\mevc1 to (111,106)\mevc1.

In the third column of Tab.~\ref{tab02} we list the production
probabilities after the capture and conversion of a {\em stopped}
$\Xi^-$ in a secondary $^9$Be target $\Xi^-+^9$Be$\rightarrow ^{10}_{\Lambda\Lambda}$Li$^*$. As before a $\Xi^-$ binding energy of 0.5\,MeV was assumed corresponding to an $^{10}_{\Lambda\Lambda}$Li excitation of about 29\,MeV.
For comparison we show in column 4 of Tab.~\ref{tab02} the results of the canonical calculations of Ref.~\cite{Yam97b} provided in their Table 5. Those calculations use a somewhat higher excitation energy of 35\,MeV.
Inspecting the excitation functions for double and twin hypernuclei
we have verified that this difference is not crucial.

While for most channels both calculations agree qualitatively, there are also important differences.  The most striking difference is seen in the production of the $^4_{\Lambda\Lambda}$H and $^5_{\Lambda\Lambda}$H double hypernuclei where the calculations differ by a factor of 4 and more than an order of magnitude, respectively. Furthermore the total produced yield for $^9_{\Lambda\Lambda}$Li is about a factor of 5 higher in our model which is mainly related to the presence of excited states. Note also, that the production of $^4_{\Lambda}$H+$^4_{\Lambda}$H twins is even at an excitation energy of 35\,MeV energetically not possible and does therefore not occur in our micro-canonical model.

As an alternative production scheme we also consider the quasi-free/{\color{black}rescattering} picture of Yamamoto {\em et al.} \cite{Yam97b} resulting in the production of excited  $^8_{\Lambda\Lambda}$He or  $^8_{\Lambda\Lambda}$H nuclei. As in ref.~\cite{Yam97b} the excitation energy of the initial  $^8_{\Lambda\Lambda}$He and  $^8_{\Lambda\Lambda}$H nuclei was fixed to 33\,MeV. The last columns of Tab.\ref{tab02} show the results from Ref.~\cite{Yam97b} while Columns 5 and 6 contain the values predicted by our micro-canonical model at the same excitation energy for this scenario. In both models the production of $^5_{\Lambda\Lambda}$H, $^6_{\Lambda\Lambda}$He and $^7_{\Lambda\Lambda}$He as well as the $^4_{\Lambda}$H+$^4_{\Lambda}$H twins dominate. Due to the higher spin factor, the production of excited states of $^4_{\Lambda}$H is enhanced over the direct ground state production in our model.

Let us first discuss the structure at (114,104){\mevc1} which has been attributed to double hypernuclei decays. Generally the production of double hypernuclei is energetically favored over the production of twins: all possible channels with
twin-production lie energetically significantly above the thresholds for
$^9_{\Lambda\Lambda}$Li, $^7_{\Lambda\Lambda}$He and
$^6_{\Lambda\Lambda}$He production in case of the $^{10}_{\Lambda\Lambda}$Li compound picture. Correspondingly the production of $^3_{\Lambda}H$ and $^6_{\Lambda}He$ twins which has been suggested \cite{Kum02} as a possible source of the peak structure around (114,104)\mevc1 is in our model by a factor of 30 and in the model of Ref. \cite{Yam97b} by a factor of 10 lower than the $^7_{\Lambda\Lambda}$He production probability.
Unlike in Ref.~\cite{Yam97b} the $^7_{\Lambda\Lambda}$He probability exceeds also the one of a $^4_{\Lambda\Lambda}$H by more than two orders of magnitude and the production of $^5_{\Lambda\Lambda}$H by a factor of about 17 in our model. Similarly, also within the quasi-free/rescattering picture the $^7_{\Lambda\Lambda}$He nucleus is the most probable channel. Of course, a direct comparison with the E906 data requires a detailed consideration of the branching ratios for pionic two-body decays many of which are not known so far.
Keeping that caveat in mind our microcanonical model supports independent on the assumed production scheme the interpretation of the E906 observation by Randeniya and Hungerford \cite{Ran07} in terms of $^7_{\Lambda\Lambda}$He decays. Decays from the ground or even exited states of $^8_{\Lambda\Lambda}$Li or $^9_{\Lambda\Lambda}$Li can possibly contribute some background to the structure at (114,104)\mevc1.

In order to describe the (133,114)\mevc1 structure of the E906 experiment, the production of $^3_{\Lambda}$H+$^4_{\Lambda}$H twins seems mandatory \cite{Ahn01}. However, the $^3_{\Lambda}$H+$^4_{\Lambda}$H+t mass lies above the initial mass $m_0= m(\Xi^-)+m(^9Be)$ and can therefore not be produced in the $\Xi^-+^9$Be compound production scheme.
With an energy of 12.6\,MeV below $m_0$, the channel $^4_{\Lambda}$H+$^6_{\Lambda}$He is the most likely twin in the present scenario, followed by the $^4_{\Lambda}$H+$^5_{\Lambda}$He+n decay. Given however the experimental precision of about 1\,\mevc1 for the momentum calibration in E906 \cite{Ahn01}, neither the decays of $^{4}_{\Lambda}$H+$^6_{\Lambda}$He with (133,108){\mevc1} nor decays of
$^{4}_{\Lambda}$H+$^5_{\Lambda}$He pairs with (133,99){\mevc1} seem to explain the structure around (133,114){\mevc1}. Of course particularly the first one could contribute to this enhancement in the tail region.

As can be seen from the last four columns of Tab.\ref{tab02} also in the quasi-free/rescattering picture the production of $^3_{\Lambda}$H+$^4_{\Lambda}$H twins plays only a minor role and none of the dominant decay channels can account for the observed structure at (133,114){\mevc1} even though
the yield ratio for  $^7_{\Lambda\Lambda}$He to $^3_{\Lambda}$H+$^4_{\Lambda}$H twin production of about 8.7 in ref.~\cite{Yam97b} resp. 5.2 in our model are still compatible with the estimated ratio mentioned above. More intriguing is however the fact that both statistical models predict a production of $^4_{\Lambda}$H+$^4_{\Lambda}$H twins even exceeding the $^3_{\Lambda}$H+$^4_{\Lambda}$H production considerably. Considering furthermore the branching ratios for two-body $\pi^-$ decays of
$\Gamma_{\pi^-+^3He}/\Gamma_{total} \approx 0.26$ \cite{Kam98} and
$\Gamma_{\pi^-+^4He}/\Gamma_{total} \approx 0.5$ \cite{Out98} the absence of a bump which could be attributed to $^4_{\Lambda}$H+$^4_{\Lambda}$H is particularly puzzling.

It seems that a process different from the ones discussed so far is required to explain the singular structure at (133,114){\mevc1} in terms of $^3_{\Lambda}$H+$^4_{\Lambda}$H twins.
Indeed in a direct mechanism like $K^- + ^9$Be${\rightarrow}$ 2n+
$K^+ + ^7_{\Lambda\Lambda}$He$^*$ the production of $^4_{\Lambda}$H+$^4_{\Lambda}$H would not be feasible and $^3_{\Lambda}$H+$^4_{\Lambda}$H pairs would be enhanced. 
Clearly for such a non-equilibrium process and considering furthermore that in the analysis of the E906 experiment e.g. cuts on the 'missing mass' of the $\Xi$ were applied \cite{Ahn01}, is difficult to predict the excitation energy distribution for the $^7_{\Lambda\Lambda}$He$^*$ nuclei. This reaction scheme is therefore at present beyond the scope of our equilibrium model and needs further detailed consideration of the initial interaction. It also is clear that the present statistical decay model needs to be complemented by quantitative weak decay calculations (see e.g. \cite{Gal}) to further substantiate this conjecture.

In summary we have presented a micro-canonical decay model to describe the break-up of an excited double hypernucleus after the absorption and conversion of a stopped $\Xi ^{-}$ hyperon. Generally the formation of excited states dominates in our model. For different $\Xi^-$ absorbing target nuclei, different produced double hypernuclei dominate. Once combined with a weak decay model these calculations will enable more reliable estimates of the $\gamma$-ray yields for the double hypernucleus measurements envisaged by the \Panda\ collaboration \cite{Poc04} and a more consistent interpretation of the E906 data. In future studies we plan to extend the present model to the production of bound $\Xi\Lambda$  or triple $\Lambda$ hypernuclei after the conversion of a $\Omega^-$ into an excited $\Xi\Lambda$ nucleus. Furthermore, the decay of excited single hyperfragments produced in electron scattering experiments will be studied with the present approach.

We thank Avraham Gal for his valuable comments and suggestions.
This work was supported by the BMBF under Contract numbers 06MZ225I and 06MZ9182. A.S.L. acknowledges the support from the State of Rhineland-Palatinate via the Research Centre `Elementary Forces and Mathematical Foundations' (EMG). We also thank the European Community-Research Infrastructure Integrating Activity "Study of Strongly Interacting Matter" (HadronPhysics2, Grant Agreement n. 227431; SPHERE network) under the Seventh Framework Programme of EU for their support.

\end{document}